\documentclass{aes2e}

\jyear{}
\jmonth{}
\jvol{}
\jnum{}

\expandafter\let\csname equation*\endcsname\relax
\expandafter\let\csname endequation*\endcsname\relax

\usepackage{amsmath}
\usepackage{graphicx}
\usepackage{booktabs}
\usepackage{musicography}
\usepackage{url}
\usepackage{amsfonts}
\usepackage{multirow}

\DeclareMathAlphabet{\mathcal}{OMS}{cmsy}{m}{n}

\newcommand{\fref}[1]{Fig.~\ref{#1}}
\newcommand{\tref}[1]{Table~\ref{#1}}
\newcommand{\sref}[1]{Sec.~\ref{#1}}
\newcommand\linesubsec[1]{\vspace{0.8mm}\noindent\textbf{#1 --- }}

\begin{document}

\markboth{STEINMETZ ET AL.}{AUDIO EFFECT STYLE TRANSFER }

\title{Style Transfer of Audio Effects with \\ Differentiable Signal Processing}


\authorgroup{
\author{CHRISTIAN J. STEINMETZ$^{1}$\thanks{\vspace{-0.0cm}{\vspace{-0.0cm}Work performed in part during an Adobe Research internship.}}} \quad \author{NICHOLAS J. BRYAN$^{2}$}  \quad
 \author{JOSHUA D. REISS$^{1}$}
\affil{$^{1}$Centre for Digital Music, Queen Mary University of London\\
$^{2}$Adobe Research, USA}
}


\abstract{
We present a framework that can impose the audio effects and production style from one recording to another by example with the goal of simplifying the audio production process.
We train a deep neural network to analyze an input recording and a style reference recording, and predict the control parameters of audio effects used to render the output.
In contrast to past work, we integrate audio effects as differentiable operators in our framework, perform backpropagation through audio effects, and optimize end-to-end using an audio-domain loss. 
We use a self-supervised training strategy enabling automatic control of audio effects without the use of any labeled or paired training data.
We survey a range of existing and new approaches for differentiable signal processing, showing how each can be integrated into our framework while discussing their trade-offs. 
We evaluate our approach on both speech and music tasks, demonstrating that our approach generalizes both to unseen recordings and even to sample rates different than those seen during training. 
Our approach produces convincing production style transfer results with the ability to transform input recordings to produced recordings, yielding audio effect control parameters that enable interpretability and user interaction.
\vspace{-0.0cm}
}

\maketitle
 

%

\section{INTRODUCTION}
\label{sec:intro}
Audio effects are commonly used in audio and video content creation to manipulate sound attributes such as loudness, timbre, spatialization, and dynamics~\cite{wilmering2020history}. 
While audio effects can be powerful tools in the hands of trained audio engineers, their complexity poses a significant challenge for less experienced users, often requiring a time-consuming adjustment process even for professionals. 
To simplify their usage, expert-designed presets can be used to configure an audio effect, but this only results in static content-independent configurations of effects.
Therefore, presets commonly require fine-tuning and adjustment, without the ability to adapt to variations across inputs.

Automatic audio production methods aim to overcome this limitation with analysis and subsequent adaptation of effects based on an input signal~\cite{IMPbook19}. These methods can provide intelligent, automatic control of a single effect parameter, multiple effects, and/or many channels of audio in a cross-adaptive manner.
Such methods may also provide a range of interfaces, from fully automatic systems to systems that only provide suggestions.
As a result, successful automatic audio production systems have the ability not only to simplify the process for amateurs, but also expedite the workflow of professionals~\cite{moffat2019approaches}.

Automatic audio production systems are commonly categorized into rule-based systems built on audio engineering best-practices, or data-driven machine learning systems~\cite{moffat2019approaches}.
While rule-based systems have seen success~\cite{deman2013b}, they are limited by the inability to formalize rules that cover the diversity arising in real-world audio engineering tasks~\cite{deman2017towards}.
Classical machine learning approaches on the other hand provide greater flexibility, but have thus far been limited by an inability to collect sufficient parametric data describing audio production in a standardized way~\cite{scott2011analysis}. Recently, deep learning approaches have shown promise in overcoming these challenges, fueling an increasing interest in data-driven audio production techniques.

Deep learning approaches for automatic audio production can be organized into direct transformation, parameter prediction, and differentiable digital signal processing methods. 
Direct transformation methods map untreated audio to a desired target and have shown promise in tasks such as speech enhancement~\cite{pascual2017segan}, source separation~\cite{stoter2019open}, and audio effect modeling~\cite{hawley2019signaltrain, martinez2020deep,wright2020real,martinez2021deep}, but have had limited use for automatic production due to a lack of interpretable control, artifacts, and/or computational complexity~\cite{martinez2021deep}. 
Parameter prediction methods predict the settings of audio effects, but require expensive ground truth parameter data~\cite{hawley2019signaltrain} or optimize over an undesirable loss in the parameter domain~\cite{sheng2019feature, mimilakis2020one}. 

Lastly, differentiable signal processing uses neural networks to control digital signal processors (DSP) implemented as differentiable operators enabling training via backpropagation~\cite{engel2020ddsp}. 
This approach imposes a stronger inductive bias by restricting the flexibility of the model, which may help to reduce the likelihood of processing artifacts, enable user control and refinement of the model predictions, and improve computational efficiency.
However, this approach requires manual implementation and often modification of DSP, imparting high engineering cost, potentially limiting its application. 
To work around this, two alternative methods have been proposed including neural proxies (NP) of audio effects~\cite{steinmetz2021automatic} and numerical gradient approximation schemes~\cite{ramirez2021differentiable}.
However, given that these approaches were proposed and evaluated in different tasks, it is difficult to fully understand their relative performance.


Beyond integrating audio effects as differentiable operators, a central limitation of deep learning systems for automatic audio production lies in the difficulty of sourcing sufficient data for supervised training, requiring both the unprocessed and final produced recordings and/or parameter data. 
In addition, the subjective and context-dependent nature of the audio production process further complicates the task~\cite{lefford2021context}. 
While evidence suggests the existence of ``best-practices''~\cite{pestana2014intelligent}, learning these techniques in a supervised paradigm is challenging.
In particular, these systems must not only learn how to control effects to achieve a desired result, they must also implicitly uncover the distribution of possible styles present within the training data across different genres and contexts.

\begin{figure}[t!]
    \centering
    \vspace{-0.0cm}
    \includegraphics[width=0.95\linewidth]{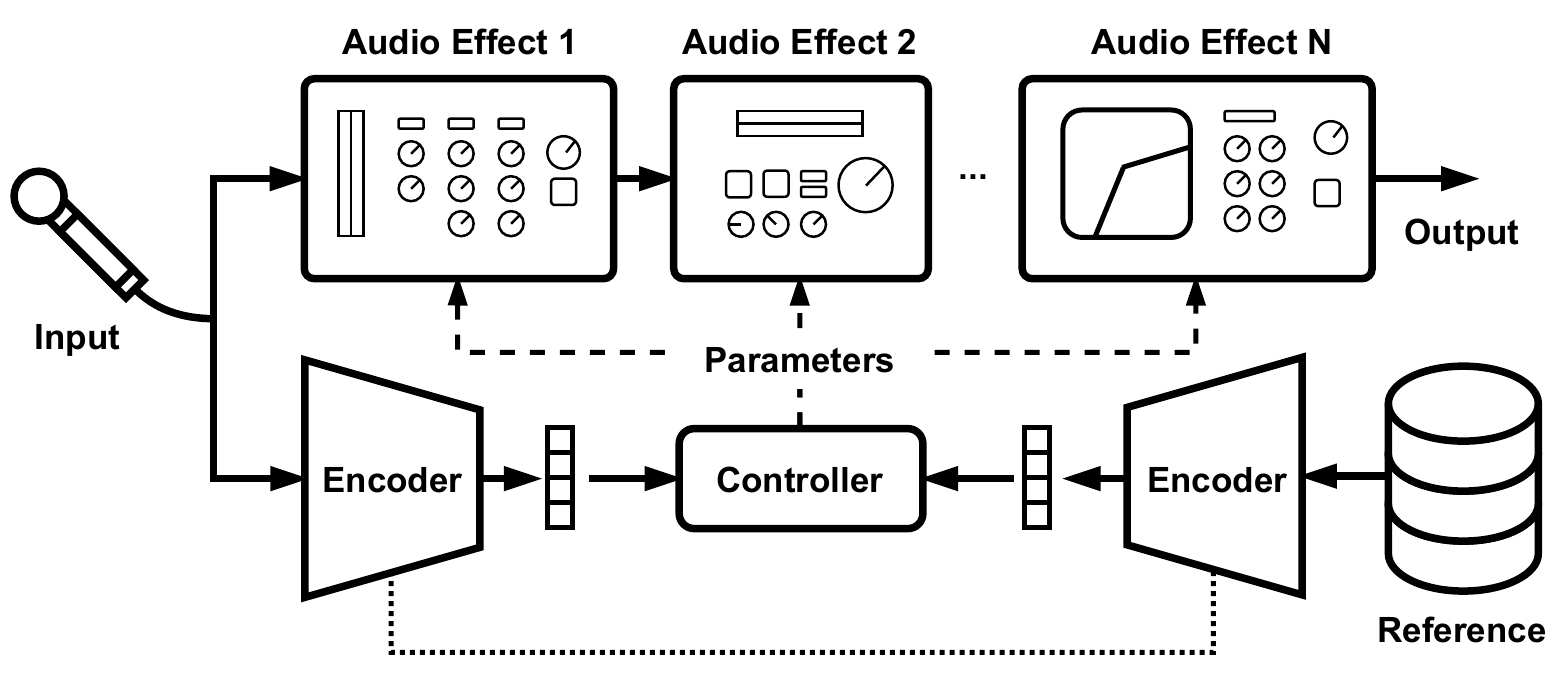}
    \vspace{-0.0cm}
    \caption{Our \emph{DeepAFx-ST} method imposes the audio effects and production style from one recording to another by example. We use a shared-weight encoder to analyze the input and a style reference signal, then compare each with a controller that outputs the parameters of effects that themselves perform style manipulation.}
    \label{fig:style-transfer-headline}
\end{figure}

In this work, we present \emph{DeepAFx-ST} -- an approach that can impose the audio effects and production style from one recording with automatic control of a set of audio effects as shown in \fref{fig:style-transfer-headline}. 
Only a short example style recording is needed at inference and no labeled training data or retraining is needed to adapt to new styles. 
To do so, we propose 
\begin{itemize}
    \setlength\itemsep{0.9em}
    \item The first audio effects style transfer method to integrate audio effects as differentiable operators, optimized end-to-end with an audio-domain loss, 
    \item Self-supervised training that enables automatic audio production without labeled or paired training data,
    \item A benchmark of five differentiation strategies for audio effects, including compute cost, engineering difficulty, and performance,
    \item The development of novel neural proxy hybrid methods, and a differentiable dynamic range compressor.
\end{itemize}

We find our approach offers several benefits including the ability to perform audio effect style transfer for both speech and music signals, yield interpretable audio effects control parameters enabling user interaction, and operate at sampling rates different than those seen during training.
Implementation of our code, a demonstration video, and listening examples are made available online\footnote{\url{https://csteinmetz1.github.io/DeepAFx-ST}}.


\section{BACKGROUND}
\label{sec:background}

\subsection{Audio Effects and Production Style Transfer}
Relevant works on audio production style transfer include initial work on controlling a dynamic range compressor with a multi-stage training procedure using neural networks and random forest~\cite{sheng2019feature}, as well as a deep neural network approach to control a parametric frequency equalizer~\cite{mimilakis2020one}. 
Both works, however, only consider controlling one audio effect in isolation and use a (quantized) parameter domain loss to avoid backpropagating through audio effects -- a non-trivial task. 
This approach has notable disadvantages including undesirable quantization schemes or loss tuning and low correlation between parameters and audio, which have been noted in the context of parameter inference for automatic synthesizer control~\cite{yee2018automatic, esling2020flow}. 
Both of these realities limit performance and generalization across classes of effects. 
However, differentiable signal processing provides the potential to overcome these challenges and has yet to be investigated for the style transfer task. 

\subsection{Differentiable Audio Effects}\label{sec:diff_fx}
Typically, differentiable signal processors are manually implemented within automatic differentiation frameworks~\cite{abadi2016tensorflow, paszke2019pytorch}.
Assuming the mathematical operations in question are differentiable, this provides a straightforward way to integrate DSP operations with neural networks. 
This approach, however, requires expert knowledge, design trade-offs, and can be difficult or impossible to implement exactly. 
Existing audio effects implemented in this manner include infinite impulse response (IIR) filters~\cite{nercessian2020neural}, reverberation~\cite{lee2021differentiable}, echo cancellers~\cite{casebeer2021auto}, DJ transitions~\cite{chen2021automatic}, and reverse engineered effects~\cite{colonel2021reverse}. 
These approaches are related to the growing body of work focused on the construction of audio synthesis models with differentiable components, which now include additive~\cite{engel2020ddsp}, subtractive~\cite{masuda2021synthesizer}, waveshaping~\cite{hayes2021neural}, and wavetable~\cite{shan2021differentiable} synthesizers. 


Differentiable audio effects of interest to our work include the parametric frequency equalizer (PEQ) and dynamic range compressor (DRC) -- two common audio production effects.
To our knowledge, there is no known past work on differentiable compressors, but differentiable PEQs do exist~\cite{nercessian2020neural}. 
PEQs are typically constructed with a cascade of multiple second-order IIR filters, also know as biquads~\cite{valimaki2016all}. 
While it is possible to implement differentiable IIR filters in the time-domain~\cite{kuznetsov2020differentiable}, the recursive filter structure can cause issues as a result of vanishing/exploding gradients and computational bottlenecks during backpropagation through time (BPTT)~\cite{hochreiter1997long}. 
This motivates frequency-domain finite impulse response (FIR) approximations~\cite{nercessian2020neural, nercessian2021lightweight, colonel2021iirnet}. In this case, the complex frequency response of the $k^{\textrm{th}}$ biquad $H_k(e^{j \omega})$ is expressed as a ratio of the DFT of the numerator and denominator coefficients, $\mathbf{b}_k = [b_{0,k}, b_{1,k}, b_{2,k}]$ and $\mathbf{a}_k = [a_{0,k}, a_{1,k}, a_{2,k}]$,
\begin{equation}
    H_k(e^{j \omega}) = \frac{ \textrm{DFT}(\mathbf{b}_k) }{ \textrm{DFT}(\mathbf{a}_k) } = \frac{\sum_{m=0}^{2} \ b_{m,k} \ e^{-j\omega m}}{\sum_{n=0}^{2} \ b_{n,k} \ e^{-j \omega n}}.
\end{equation}
The overall response of a cascade of $K$ biquad filters is then given by the product of their responses $H_{sys} (e^{j \omega}) = \prod_{k=1}^{K} H_k(e^{j \omega})$.
We can approximate the response using an FIR filter by evaluating $H_{sys}(e^{j \omega})$ over a linearly spaced, zero-padded frequencies of length $2^{\lceil \log_2 (2N - 1) \rceil}$, applying the filter(s) to an input $x[n]$ with $N$ samples in the frequency domain by multiplying with  the input $ X(e^{j \omega})$, 
and then performing the inverse DFT to get the output $y[n]$.


\subsection{Alternative Differentiation Methods}

In contrast to manually implementing differentiable signal processing operations, three alternative approaches are of note. 
In neural proxy (NP) approaches, a neural network is trained to emulate the behavior of a signal processor~\cite{grathwohl2017backpropagation, jacovi2019neural} including control parameter behavior using a dataset of input/output measurements. 
While most work in this area has focused on creating emulations of analog audio effects~\cite{hawley2019signaltrain, martinez2020deep, wright2020real, martinez2021deep, steinmetz2021efficient, damskagg2019deep, martinez2019modeling, wright2019real}, this approach can be used together with a larger neural network for automatic control of effects, as was demonstrated with applications in automatic mixing~\cite{steinmetz2021automatic}.
Since neural proxy models are composed of neural network building blocks, they are differentiable by default, but require accurate emulation to work effectively and can be computationally complex.  

In an effort to reduce inference time complexity and reduce approximation error, neural proxy models can be used together with the original DSP device to create what we call neural proxy hybrid models~\cite{tseng2019hyperparameter}.
Half hybrid approaches (NP-HH) use the neural proxy during the forward and backward pass of training, but use the DSP device during inference.
Full hybrid approaches (NP-FH) aim to further reduce approximation error caused by the neural proxy by using the DSP device during inference as well as the forward pass during training.
In this case, the neural proxy is used only during the backward pass where gradient computation is required.
Thus far neural proxy hybrid approaches, however, have not been applied in audio, and have only been used in computer vision for hyperparameter optimization, and not in an adaptive manner for automatic control.

As a third alternative, non-differentiable DSP implementations can be used directly with numerical gradient approximation methods, which has been demonstrated in audio effect modeling, removal of breaths, and music mastering~\cite{ramirez2021differentiable}. 
In context of controlling audio effects $h(x, p)$ with input audio $x$ and parameters $p \in \mathbb{R}^{P}$, we only need to estimate the partial derivatives for each parameter $\frac{\partial}{\partial p_i} h(x,p)$. 
For this, finite differences (FD)~\cite{milne2000calculus} perturb each input parameter forward and backward and evaluate the operator $2P$ times for $P$ control parameters. 
Alternatively, simultaneous permutation stochastic approximation (SPSA)~\cite{spall1998overview} gradient estimation only requires two function evaluations per gradient vector estimate. 
In this case, the approximate partial derivative with respect to $p_i$ is given by 
\begin{equation}
\frac{\hat{h}(x, p_i)}{p_i} = \frac{h(x, p + \epsilon \Delta^P) - h(x, p - \epsilon  \Delta^P)}{2 \epsilon \Delta^{P}_{i}},
\end{equation}
where $\epsilon$ is a small, non-zero value and $\Delta^P \in \mathbb{R}^P$ is a random vector sampled from a symmetric Bernoulli distribution ($\Delta^P_i = \pm 1$)~\cite{spall1992multivariate}.  
This approach does not require pre-training or knowledge of the DSP, but can be challenging to train due to inaccuracies in gradient approximation and overhead from implementations available only on CPU.

\begin{figure*}[t!]
    \centering
    \vspace{-0.0cm}
    \includegraphics[width=0.99\linewidth,trim={0cm 0.1cm 0.8cm 0},clip]{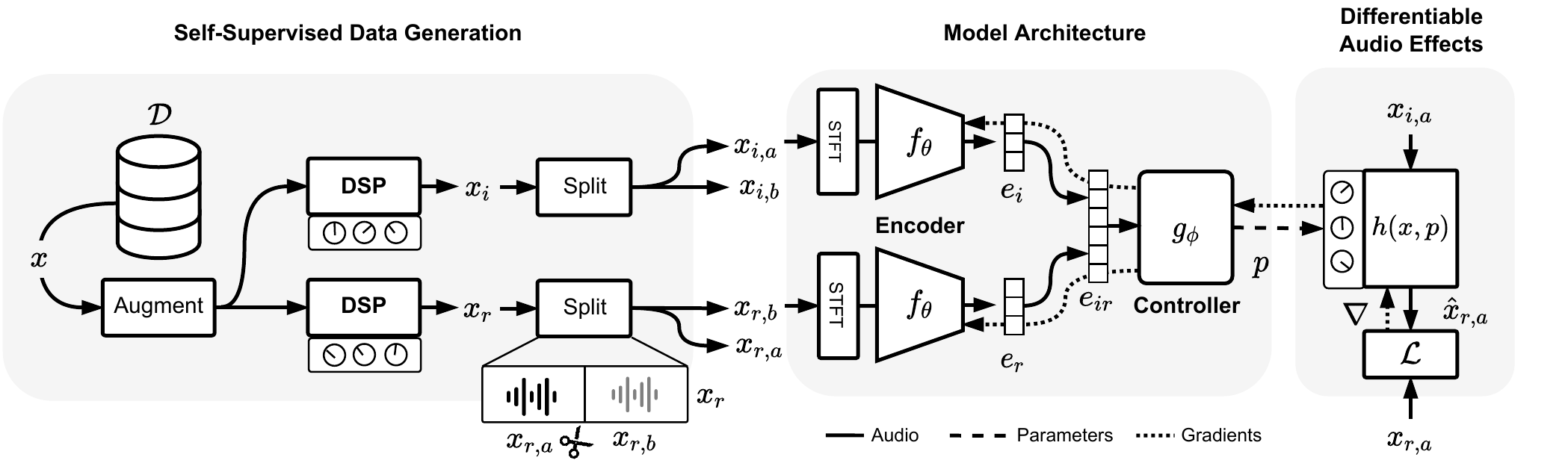}
    \vspace{-0.0cm}
    \caption{The self-supervised training process for our \emph{DeepAFx-ST} approach.
    We sample $x$ from a large dataset of recordings $\mathcal{D}$ and apply random augmentations.
    Recordings with two different styles are produced with randomly configured DSP, which become the input $x_i$ and the style reference $x_r$.
    We split these recordings in half, and pass the opposite sections ($x_{i,a}$ and $x_{r,b}$) to shared-weight encoders $f_\theta$. 
    Using the concatenated encoder representation $e_{ir}$, the controller $g_\phi$ estimates parameters $p$ to apply the reference style to the input.}
    \label{fig:data-generation}
    \vspace{-0.0cm}
\end{figure*}

\section{PROPOSED METHOD}
\label{sec:method}
Our proposed audio production style transfer method is composed of three components as shown in \fref{fig:data-generation}.
A self-supervised data generation process enables the creation of input and reference recordings, $x_i$ and $x_r$, from a large dataset of audio $\mathcal{D}$.
These input and reference recordings are generated by processing a source recording through two different configurations of a chain of audio effects, resulting in two different production styles.  
Then a shared weight encoder $f_\theta$ is used to extract information about the production style of the input and reference recordings. 
This information is aggregated and then passed to the controller network $g_\phi$ that is tasked with estimating parameters $p$ to configure the differentiable audio effects $h(x,p)$ so that the processed input signal $\hat{x}_r$ matches the style reference $x_r$.

\subsection{Self-Supervised Training}
\label{sec:ssl}

Our self-supervised training strategy enables learning control of audio effects without labeled or paired training data. 
Our data generation process is shown on the left of~\fref{fig:data-generation} and begins by selecting a random audio recording $x$ from a dataset $\mathcal{D}$. 
We then create an augmented version $x_{aug}$ by applying pitch shifting and time stretching to further increase diversity.
Two versions of this augmented recording are created, $x_i$ and $x_r$, each with a different production style. 
The recording in the top branch $x_i$ becomes the input, and the recording in the bottom branch $x_r$ becomes the style reference.
This is achieved by applying a set of audio effects to both recordings, with each set of effects having a different random configuration.
To ensure that sensible outputs are produced, the method for sampling random effect configuration should be carefully tuned, as is common with data augmentation pipelines. 

To train our model to only focus on audio production style and operate when the content differs between the input and style reference, we split the recordings $x_i$ and $x_r$ in half, generating an $a$ and $b$ section, producing four recordings $x_{i,a}$, $x_{i,b}$, $x_{r,a}$, and $x_{r,b}$.
During training, we randomly select either the $a$ or $b$ section to be passed as input and use the other section as the style reference.
To compute an audio-domain loss we use the style reference corresponding to the same section as the input in order to have a paired input and ground truth with the same content. 
Typical approaches to constructing audio datasets often require ground truth paired recordings, or generate them by taking clean recording and degrading them.
However, our approach generates paired data using a dataset of any source material with a range of quality and recording conditions. 

The process of splitting the recordings does not perfectly mimic the case at inference, where the input and reference recordings come from different sources.
However, this splitting operation helps to further increase the difference between the input and reference recordings during training, as opposed to using recordings with the same content.
The process of randomly selecting either the $a$ or $b$ section as input and reference further increases this diversity by ensuring that the model cannot rely on the relative temporal relation of the input and style reference.

\subsection{Model Architecture}
\label{sec:arch}

To carry about the production style transfer task, magnitude spectrograms from the input and style reference recordings are fed to a shared-weight convolutional neural network encoder $f_\theta$, which produces a time series of embeddings for each recording.
Temporal average pooling is used to aggregate these embeddings across time, producing a single embedding of dimension $D$ for both the input $e_i$ and style reference $e_r$. 
These embeddings are then concatenated $e_{ir}$ and passed to the controller network.

The controller network $g_\phi$ is a basic multi-layer perception (MLP) and is tasked with producing control parameters $p$ to configure a set of audio effects $h(x, p)$.
This network should consider the information from the encoder about the production style of the input and the reference.
The estimated control parameters should configure the set of audio effects in such a way that passing the input signal though the effect chain will produce a recording that matches the style reference. 

\subsection{Differentiable Audio Effects}
\label{sec:diff_methods}
Core to our approach is the integration of audio effects directly within the computation graph of a neural network. 
Learning to control audio effects instead of carrying out audio processing with a neural network helps us incorporate signal processing domain-knowledge, impose a strong inductive bias, reduce processing artifacts, and reduce computational complexity~\cite{engel2020ddsp}. 
In contrast to past work~\cite{mimilakis2020one, sheng2019feature}, we fully integrate audio effects as differentiable operators or layers with standard automatic differentiation tools and enable backpropagation through effects during training as shown with dashed lines in Figure~\ref{fig:data-generation}. 

We propose backpropagating through audio effects using five unique differentiation strategies.
These include manually implemented automatic differentiation effects (AD), neural proxy effects (NP), full neural proxy hybrids (NP-FH), half neural proxy hybrids (NP-HH), and numerical gradient approximations (SPSA).
While AD, NP, and SPSA methods have previously been employed in automatic audio production tasks, NP-FH has only been used in static image processing hyperparameter optimization~\cite{tseng2019hyperparameter}, and NP-HH has not been proposed before. 
Moreover, all approaches have never been compared in a unifying manner, leaving uncertainty around their relative efficacy.

\section{EXPERIMENTAL DESIGN}
\label{sec:experimental_design}

\begin{figure}[t!]
    \centering
    \includegraphics[width=\linewidth, trim={0.0cm 0.18cm 0.0cm 0.1cm},clip ]{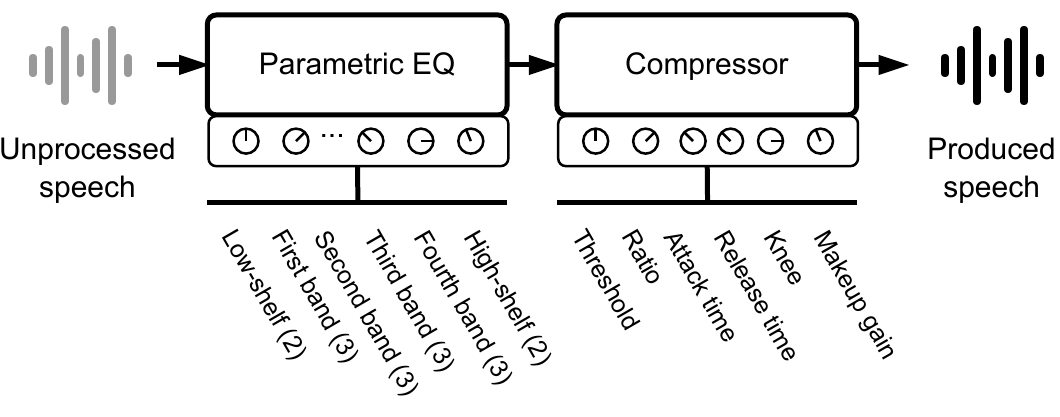}
    \caption{Our audio effect signal chain consisting of a 6-band parametric equalizer and single-band dynamic range compressor.}
    \label{fig:signal-chain}
    \vspace{-0.0cm}
\end{figure}

To demonstrate our approach, we focus on the task of imposing the equalization and loudness dynamics from one recording to another using a parametric equalizer (PEQ) and dynamic range compressor (DRC) as shown in~\fref{fig:signal-chain}. 
This forms the basic signal chain employed by audio engineers with applications in dialogue processing or in music mastering.
We compare performance of our production style transfer system using each of the differentiation strategies introduced in ~\sref{sec:diff_methods}, keeping the data generation and network architecture fixed.
By doing so, we seek to both understand the performance of our production style transfer approach, as well as the advantages and disadvantages of these differentiation strategies.

\subsection{Differentiable Equalizer and Compressor}
\label{sec:eq_drc}
We implement a differentiable PEQ following past work~\cite{nercessian2020neural}, but develop our own novel differentiable dynamic range compressor. 
For our differentiable compressor, we use a traditional feedforward dynamic range compressor, which is in fact composed of differentiable operations~\cite{giannoulis2012digital}.
Similar to PEQs, however, we found direct implementation of existing compressor designs problematic due to the presence of recursive filters, in this case due to the smoothing filters for the ballistics,
\begin{equation} \label{eq:smoothing}
    \resizebox{.9\hsize}{!}{
   $y_L[n] =
    \begin{cases}
        \alpha_A y_L[n-1] + (1-\alpha_A) x_L[n] & x_L[n] > y_L[n-1] \\
        \alpha_R y_L[n-1] + (1-\alpha_R) x_L[n] & x_L[n] \leq y_L[n-1]
    \end{cases}$,       
    }
\end{equation}
where $y_L[n]$ is the smoothed gain reduction, $x_L[n] = x_G[n] - y_G[n]$ is the result of subtracting the output of the gain computer from the signal envelope, and $\alpha_A$ and $\alpha_R$ are the attack and release time constants. 
This operation is not easily approximated with a FIR filter, thus we simplify our design using a single attack and release time constant $\alpha$ to create a single-pole IIR ballistics filter,
\begin{equation}
   y_L[n] = \alpha y_L[n-1] + (1-\alpha) x_L[n].
\end{equation}
This filter can be approximated using an FIR filter by employing the same approach as in differentiable PEQs.
While this modification restricts the capabilities of our compressor, forcing the effective attack and release times to be shared, we seek only a basic compressor to show the complexity in implementing differentiable versions of common audio effects, which we compare against competing alternative differentiation strategies. 

\subsection{Datasets}
\label{sec:datasets}

We employ our data generation pipeline as described in~\sref{sec:method} using both speech and music.
To test our approach on speech, we use \texttt{train-clean-360}, a subset of LibriTTS~\cite{zen2019libritts}, which totals 360 hours of audio at $f_s = 24$\,kHz.
To test our approach on music, we use the MTG-Jamendo dataset~\cite{bogdanov2019mtg}, which provides over 55\,k songs, which we downmix to mono and resample to $f_s = 24$\,kHz. 
For both datasets, we generate a $90$/$5$/$5$ split of the recordings for training, validation, and testing. When randomly sampling patches of audio from a recording, if the selected segment $x[n]$ has a mean energy below the threshold, $\frac{1}{N} \sum_{n=1}^{N} | x[n] |^2 \leq 0.001$, we continue sampling until a non-silent segment is found.
To ensure that there is sufficient headroom for processing within the audio effects, we peak normalize all inputs and style recordings to $-12$\,dBFS.

\subsection{Baselines}
\label{sec:baselines}

\linesubsec{Rule-based DSP}
\label{sec:dsp}
Our rule-based baseline consists of an automatic, two stage process that aims to mimic the signal chain of the PEQ and DRC. 
First, an FIR filtering stage is employed, which aims to transform the spectral content of the input to match the spectrum of the style reference. 
We compute the average magnitude spectrum of the input and style reference by computing the STFT using a large window of $N=65536$ and a hop size of $H=16384$, averaging across the frames, yielding $X_i$ and $X_r$, which
are smoothed using a Savitzky-Golay filter~\cite{savitzky1964smoothing}.
We design an inverse FIR filter with $63$ taps using the window method, with a desired magnitude response given by $X = X_{i,sm} / X_{r,sm}$, where $X_{i,sm}$ and $X_{r,sm}$ are the smoothed average spectra of the input and the style reference.

The filtered signal is sent to a dynamic range compressor where all parameters except for the threshold are fixed. 
In order to adjust the amount of compression based on the input signal and the style reference, the threshold starts at 0 dB and is lowered in increments of 0.5 dB.
At each iteration the loudness in dB LUFS~\cite{steinmetz2021pyloudnorm} of the output is measured and compared to the loudness of the style reference.
If the difference in the loudness measurements are below a given threshold (e.g. 0.5 dB LUFS), the process halts. 
Otherwise, the process continues until the loudness is within a specific tolerance, or the threshold reaches -80 dB.

\linesubsec{End-to-end neural network}
\label{sec:e2e}
As a baseline on the opposite end of the spectrum, we evaluate a direct, end-to-end neural network approach to replace the differentiable signal processing audio effects. 
This network utilizes the same architecture as our neural proxy approaches, but does not use pre-trained weights. 
Instead, the weights of this model are optimized during the training of the encoder and controller. 
The controller still predicts parameters for this network, as in the case of a neural proxy, since the weights of the network are not frozen during training.
As a result the model will learn its own set of control parameters. 
We evaluate two variants of this approach, first using a single TCN network (cTCN 1) and a second using two TCNs connected in series (cTCN 2), which mimics the setup in the neural proxy approach.
\subsection{Losses}
We train both our neural proxy models and production style transfer models using an audio-domain loss.
In particular, we compute our loss using the model output and ground truth audio from our data generation process with a weighted sum of the error in time and frequency domains.
The time domain loss $\mathcal{L}_{\text{time}}$ is the mean absolute error (MAE) and the
frequency domain loss $\mathcal{L}_{\text{freq}}$ is the multi-resolution short-time Fourier transform loss (MR-STFT)~\cite{wang2019neural, steinmetz2020auraloss}.
This loss is the sum of the $L_1$ distance between the STFT of the ground truth and estimated waveforms measured both in the $\log$ and linear domains, at multiple resolutions, in this case with window sizes $W \in [32, 128, 512, 2048, 8192, 32768]$ and hop sizes $H = W/2$.
The overall loss is  $\mathcal{L}_{\text{overall}}({\hat{y}},y) = \mathcal{L}_{\text{freq}}({\hat{y}},y) + \alpha \cdot \mathcal{L}_{\text{time}}({\hat{y}},y) \label{eq:overall-loss}$, where $\alpha = 100$.

\setlength{\tabcolsep}{3.0pt}
\setlength{\abovecaptionskip}{15pt plus 3pt minus 2pt}
\begin{table*}[t!]
    \centering
    \resizebox{\textwidth}{!}{%
    \begin{tabular}{l c c c c c c c c c c c c c c c c c c c c} \toprule
        \multicolumn{21}{c}{\normalsize{\textbf{Speech}}} \\ \midrule
        & \multicolumn{6}{c}{\textbf{LibriTTS}} & & \multicolumn{6}{c}{\textbf{DAPS}} & & \multicolumn{6}{c}{\textbf{VCTK}} \\ \cmidrule(lr){2-7} \cmidrule(lr){8-14} \cmidrule(lr){15-21}
        \textbf{Method}  &  \footnotesize{\textbf{PESQ}} &  \footnotesize{\textbf{STFT}}  & \footnotesize{\textbf{MSD}} & \footnotesize{\textbf{SCE}} & \footnotesize{\textbf{RMS}} & \footnotesize{\textbf{LUFS}} & & \footnotesize{\textbf{PESQ}} &  \footnotesize{\textbf{STFT}}  & \footnotesize{\textbf{MSD}} & \footnotesize{\textbf{SCE}} & \footnotesize{\textbf{RMS}} & \footnotesize{\textbf{LUFS}} & & \footnotesize{\textbf{PESQ}} &  \footnotesize{\textbf{STFT}}  & \footnotesize{\textbf{MSD}} & \footnotesize{\textbf{SCE}} & \footnotesize{\textbf{RMS}} & \footnotesize{\textbf{LUFS}}\\ \midrule
        {Input}   
        & 3.765           & 1.187           & 2.180           & 687.5           & 6.983           & 2.426           & 
        & 3.684           & 1.179           & 2.151           & 641.7            & 6.900          & 2.314          & 
        & 3.672           & 1.254           &          2.008  & 815.4           & 7.783           & 2.532\\ 
        {RB-DSP}  
        & 3.856           & 0.943           & 1.955           & 410.3           & 4.204           & 1.674           & 
        & 3.787           & 0.917           & 1.882           & 399.7            & 3.705           & 1.481          & 
        & 3.709           & 1.101           &          1.911  & 657.6           & 5.039           & 2.018\\ \midrule
        {cTCN 1}    
        & 4.258           & 0.405           & 0.887           & 128.4           & 2.237           & 1.066           & 
        & 4.185           & 0.419           & 0.884           & 124.6            & 2.098           & 1.006          & 
        & 4.181           & 0.467           &          0.891  & 173.8           & 2.651           & 1.165\\
        {cTCN 2}    
        & 4.281           & 0.372           & \textbf{0.833}  & 117.5           & 1.927           & 0.925           & 
        & \textbf{4.224}  & \textbf{0.391}  & \textbf{0.841}  & 113.9            & 1.886           & 0.913          & 
        & 4.201           & \textbf{0.441}  & \textbf{0.856}  & 163.8           & 2.431           & 1.086\\ \midrule
        {NP}   
        & 3.643           & 0.676           & 1.405           & 265.0           & 2.812           & 1.340           & 
        & 3.605           & 0.685           & 1.362           & 249.2            & 2.732           & 1.350          & 
        & 3.651           & 0.737           &          1.300  & 321.7           & 3.166           & 1.453\\
        {NP-HH}   
        & 3.999           & 1.038           & 2.179           & 440.2           & 5.472           & 2.679           & 
        & 3.903           & 1.022           & 2.113           & 451.9            & 5.104           & 2.535          & 
        & 3.951           & 1.044           &          1.930  & 591.5           & 5.194           & 2.651\\
        {NP-FH}   
        & 3.945           & 1.058           & 2.088           & 404.9           & 6.820           & 3.197           & 
        & 3.891           & 1.037           & 2.045           & 395.4            & 6.754           & 3.117          & 
        & 3.894           & 1.087           &          1.934  & 514.0           & 7.065           & 3.363\\ 
       
        {SPSA}      
        & 4.180           & 0.635           & 1.406           & 219.5           & 3.263           & 1.600           & 
        & 4.099           & 0.645           & 1.379           & 213.6            & 2.989           & 1.511          & 
        & 4.023           & 0.730           &          1.359  & 301.6           & 3.535           & 1.737\\ 
        \midrule
        {AD}  
        & \textbf{4.310}  & \textbf{0.388}  & 0.882           & \textbf{111.5}  & \textbf{1.828}  & \textbf{0.823}  & 
        & 4.222           & 0.416           & 0.895           & \textbf{109.0}   & \textbf{1.758}  & \textbf{0.799} & 
        & \textbf{4.218}  & 0.481           &          0.924  & \textbf{152.7}  & \textbf{2.317}  & \textbf{1.006} \\ 
         \bottomrule 
    \end{tabular}}
    \vspace{0.1cm}
\caption{Synthetic production style transfer with models trained using LibriTTS. Held-out speakers from the LibriTTS dataset are used, while utterances from DAPS and VCTK come from datasets never seen during training. Lower is better for all metrics except PESQ.}
    \vspace{-0.1cm}
    \label{tab:style-transfer-synthetic-speech}
\end{table*}

\subsection{Neural Proxy Pre-training}
To construct our neural proxy-based models, we utilized a temporal convolutional network (TCN)~\cite{bai2018empirical}, also known as the feedforward WaveNet~\cite{rethage2018wavenet}. This model has seen success in modeling distortion~\cite{wright2019real} and compressors~\cite{steinmetz2021automatic}.
In particular, we use a causal TCN~\cite{steinmetz2021automatic} with feature-wise linear modulation (FiLM)~\cite{perez2018film} to adapt to the effect control parameters.
Our TCN configuration has $4$ blocks, kernel size of $13$, channel width of $64$ and a dilation factor that grows as a power of $8$, where the dilation factor at block $b$ is $d_b = [ 1, 8, 64, 512 ]$.
We trained a total of 4 proxies, one for each of the two effects, with one for both speech and music. 
Training examples were generated by using DSP effects with uniformly random parameters. 
We trained each model for 400 epochs, where one epoch was defined as 20,000 samples, and used the checkpoint with the lowest validation loss.
We used an initial learning rate of $3\cdot 10^{-4}$, a batch size of 16, and inputs $\approx 3$s ($65536$ samples) in length. 
The learning rate was lowered by a factor of 10 after the validation loss had not improved for 10 epochs. 

\subsection{Production Style Transfer Training}
After training our neural proxy models, we trained our production style transfer model using five different audio effects differentiation methods -- AD, NP, NP-HH, NP-FH, and SPSA as well as end-to-end cTCN1 and cTCN2. 
The encoder operates on spectrograms using a short-time Fourier transform with a Hann window of size $N=4096$ and a hop size of $H=2048$.
The magnitude of the spectrogram was computed and exponential compression was applied following $|X|^{0.3}$. 
The compressed magnitude spectrogram was then passed into a randomly initialized EfficientNet B2 CNN~\cite{tan2019efficientnet} with a single input channel. 
The final activations were projected to $D=1024$ with a linear layer, followed by temporal mean pooling to produce a single representation, which was normalized by the $L_2$ norm.

The controller network was implemented as a simple three layer MLP, along with PReLU~\cite{he2015delving} activation functions. 
At the output of the controller, the hidden representation was projected into the audio effect control parameter space $\mathbb{R}^P$, and a sigmoid activation was applied. 
All parameters were then denormalized following predefined minimum and maximum ranges for parameters before they were passed to the audio effects. 

We trained all models for 400 epochs with a batch size of 6, where a single epoch is defined as 20,000 segments of audio $\approx 10$ seconds in length ($262144$ samples). 
We used a learning rate of $10^{-4}$ for all models, with the exception of the SPSA models, which required a lower learning rate of $10^{-5}$ for stability. 
In both cases, we scheduled the learning rate such that it is lowered by a factor of 10 at both $80\%$ and $95\%$ through training.
We applied gradient clipping with a value of $4.0$ and used an SPSA $\epsilon$ of $5 \cdot 10^{-4}$.



\setlength{\tabcolsep}{2.5pt}
\begin{table*}[ht]
    \vspace{-0.2cm}
    \centering
    \resizebox{\linewidth{}}{!}{%
    \begin{tabular}{l c c c c c c c c c c c c c c c c c c c c} \toprule
    \multicolumn{21}{c}{\normalsize{\textbf{Music}}} \\ \midrule
        &  
        \multicolumn{6}{c}{\textbf{MTG-Jamendo (24 kHz)}}   & & 
        \multicolumn{6}{c}{\textbf{MTG-Jamendo (44.1 kHz)}} & & 
        \multicolumn{6}{c}{\textbf{MUSDB18 (44.1 kHz)}} \\ 
        \cmidrule(lr){2-7} \cmidrule(lr){9-14} \cmidrule(lr){15-21}
        \textbf{Method}   & 
        \textbf{PESQ} & \textbf{STFT} & \textbf{MSD} & \textbf{SCE} & \textbf{RMS} & \textbf{LUFS} & &
        \textbf{PESQ} & \textbf{STFT} & \textbf{MSD} & \textbf{SCE} & \textbf{RMS} & \textbf{LUFS} & &
        \textbf{PESQ} & \textbf{STFT} & \textbf{MSD} & \textbf{SCE} & \textbf{RMS} & \textbf{LUFS} \\ \midrule
        Input      
        & 2.927  & 1.198  & 6.088  & 646.5  & 6.695  & 2.518 & 
        & 2.874  & 1.109  & 4.454  & 767.7  & 6.793  & inf   & 
        & 2.900  & 1.252  & 4.342  & 1088.3    & 5.940  & 2.312    \\ 
        RB-DSP     
        & 2.849  & 0.925  & 4.422  & 263.2  & 4.254  & 1.706  & 
        & 2.931  & 0.887  & 3.355  & 342.0   & 4.749  & 1.882 & 
        & 2.994  & 0.821  & 3.052  & \textbf{379.4}   & 4.078 & 1.665  \\ 
        \midrule
        cTCN 1       
        & \textbf{3.402}  & 0.547  & 2.294  & 160.9  & 3.261  & 1.483 &
        & 3.168  & 0.876  & 2.921  & 494.4    & 4.372  & 2.070  & 
        & 3.121  & 0.896  & 2.956  & 730.4    & 4.548  & 2.231  \\ 
        cTCN 2       
        & 3.390  & 0.548  & 2.278  & 152.3  & 2.951  & 1.397 & 
        &  3.123  & 0.903  & 2.973  & 517.4   & 4.084  & 1.899  &
        &  3.107  & 0.917  & 2.986  & 749.5   & 4.208  & 2.061 \\  \midrule
        NP S  
        & 2.926  & 0.787  & 2.838  & 221.3  & 2.785  & 1.390 &  
        & 2.764  & 1.033  & 2.869  & 488.3  & 3.710  & 1.824 & 
        & 2.804  & 0.950  & 2.778  & 742.1   & 3.835  & 1.921  \\
        NP M  
        & 2.765  & 0.845  & 3.211  & 255.2  & 3.227  & 1.608 & 
        & 2.699  & 1.042  & 2.928  & 497.0    & 3.942  & 1.961   & 
        & 2.791  & 0.946  & 2.800  & 757.1   & 4.209  & 2.127    \\
        NP-HH (S)  
        & 2.819  & 1.092  & 6.791  & 395.3  & 6.276  & 3.032 & 
        & 2.865  & 1.101  & 5.194  & 689.2    & 6.792  & 3.365 & 
        & 2.853  & 1.165  & 4.852  & 1005.7   & 6.451  & 3.269 \\
        NP-HH (M)  
        & 2.532  & 1.166  & 7.070  & 591.9  & 5.660  & 2.593 & 
        & 2.512  & 1.148  & 5.385  & 854.9   & 5.940  & 2.740 & 
        & 2.493  & 1.198  & 5.090  & 1021.9    & 5.585  & 2.731 \\
        NP-FH (S)  
        & 2.833  & 1.016  & 5.005  & 280.8  & 3.377  & 1.634 & 
        & 2.888  & 0.977  & 3.883  & 429.9    & 3.480  & 1.709   & 
        & 2.857  & 1.045  & 3.809  & 617.3   & 3.932  & 1.971   \\
        NP-FH (M)  
        & 2.648  & 1.137  & 6.368  & 605.6  & 5.903  & 2.587 & 
        & 2.625  & 1.133  & 5.512  & 844.6   & 6.417  & 2.876  & 
        & 2.575  & 1.225  & 5.450  & 1172.5   & 5.973  & 2.913   \\
        SPSA         
        & 3.173  & 0.716  & 3.024  & 210.1  & 2.809  & 1.344 & 
        & 3.172  & 0.759  & 2.458  & 386.2   & 2.839  & 1.311 & 
        &  3.126  & 0.789  & 2.321  & 574.4   & 2.925  & 1.394    \\ 
        \midrule
        AD     
        & 3.355  & \textbf{0.488}  & \textbf{2.149}  & \textbf{144.7}  & \textbf{2.167}  & \textbf{1.005}   & 
        &  \textbf{3.400}  & \textbf{0.585}  & \textbf{1.824}  & \textbf{304.4}    & \textbf{2.425}  & \textbf{1.106} & 
        &  \textbf{3.396}  & \textbf{0.608}  & \textbf{1.695}  & 456.1    & \textbf{2.559}  & \textbf{1.197}  \\
         \bottomrule 
    \end{tabular}}
    \vspace{0.2cm}
    \caption{Synthetic production style transfer using the MTG-Jamendo dataset. Proxy models (S) were pre-trained using speech from LibriTTS and (M) were trained using music from the MTG-Jamendo dataset. Lower is better for all metrics except PESQ.}
    \label{tab:style-transfer-synthetic-music}
    \vspace{-0.2cm}
\end{table*}

\section{RESULTS}
\label{sec:results}


\subsection{Synthetic Production Style Transfer}
We first evaluated on a synthetic production style transfer task mirroring our training setup. 
This facilitates the use of full-reference metrics, since we synthetically generate style references and as a result have inputs and ground truth with matching content.
Results of the evaluation using held-out speakers from LibriTTS are shown in \tref{tab:style-transfer-synthetic-speech}.

As there are no established metrics for evaluating audio production style, we propose a range of metrics. 
For a general notion of perceptual similarity we use PESQ~\cite{rix2001perceptual}, along with the Multi-resolution STFT~\cite{wang2019neural} error, which is a component of the training objective. 
To evaluate the ability of the system to address equalizer-related style components, we report the mel-spectral distance (MSD), which consists of the error between melspectrograms using a large window size of $W=65536$, along with the spectral centroid error (SCE). 
To evaluate aspects related to the dynamic range, we consider the RMS energy error (RMS) and the perceptual loudness error (LUFS)~\cite{steinmetz2021pyloudnorm}. 

We note our rule-based DSP baseline improves on the input across all metrics, though not by a significant margin.
Overall, we find strong performance from the AD method, which performs best on all metrics except MSD, which is only slightly outperformed by cTCN2.
Both cTCN approaches follow closely behind AD, demonstrating the strength of end-to-end neural network approaches. 
The SPSA model follows closely behind these approaches with somewhat degraded PESQ and STFT error, yet it outperforms all NP methods. 
While the NP methods are not as competitive, both NP-HH and NP-FH outperform NP in terms of PESQ, likely a result of using a DSP implementation, which imparts less artifacts.
However, the NP approach achieves better performance on all other metrics, indicating it likely better matches the style reference. 

In addition to evaluating on held-out speech from LibriTTS, we also evaluate generalization with speech datasets not seen during training including DAPS~\cite{mysore2014can}, which contains high quality recordings from 20 speakers, and the larger VCTK~\cite{yamagishi2019vctk} dataset, which contains 110 speakers.
We observe a similar trend in performance, yet the cTCN models slightly outperform AD as compared to on LibriTTS.
This indicates our self-supervised training was sufficient for producing models that generalize to speech from different sources and acoustic conditions. 

We also trained the same set of models using music from the MTG-Jamendo dataset. 
The result, shown in \tref{tab:style-transfer-synthetic-music}, largely follows a similar trend to the speech experiments with regards to the differentiation methods, demonstrating the general nature of our approach. 
However, it appears that the music task is somewhat more difficult.
We also compared neural proxies trained on speech (S) or music (M), and found those trained on music performed slightly worse.

We then investigated operation at sample rates different than training. 
To achieve variable sample rate operation we resampled the signals passed to the encoder to match training (24\,kHz), however, the audio effects operate at the sample rate of the input audio (44.1\,kHz).
First, we evaluated on the MTG-Jamendo test set, and we noticed a significant drop in performance for the neural network-based approaches, which include the cTCNs and NP, since these models operate at a fixed sample rate. 
However, the DSP-based approaches performed significantly better, including both the AD and SPSA approaches, which performed only slightly worse than at the training sample rate.

Finally, to test out-of-distribution generalization, we used music from MUSDB18~\cite{musdb18}.
While performance does decrease across most metrics and for all models, the models can still perform the task, with SPSA and AD approaches exceeding the baseline. 

\subsection{Realistic Production Style Transfer}
\setlength{\tabcolsep}{3.0pt}

To simulate a more realistic production style transfer scenario where the style reference contains not only different content but also differing sources (e.g. speaker identity or instrumentation), we construct a realistic production style transfer evaluation task.
We designed five different audio production ``styles'', which are defined by hand-crafted parameter ranges inspired by commonly used plugin presets.
We generate audio from these five different styles (Telephone, Warm, Bright, Neutral, and Broadcast) by randomly sampling parametric equalizer and compressor parameters from these ranges such that each style is audibly unique.
Sampling parameter values randomly provides more variation within audio examples of the same style, which is more complex than applying a fixed preset. 

To generate examples in each of these styles, we use speech from DAPS, and music from MUSDB18.
However, since we do not have paired ground truth as in the synthetic production style transfer case, we performed our evaluation using only non-intrusive metrics that do not require a matching reference. 
These metrics measure the error between high-level audio features, like the MSD, spectral centroid, RMS energy, and perceptual loudness.

\begin{table}[t!]
    \centering
    \resizebox{1.0\hsize}{!}{%
    \begin{tabular}{l c c c c c c c c c } \toprule
        & \multicolumn{4}{c}{\textbf{DAPS}} & & \multicolumn{4}{c}{\textbf{MUSDB18}} \\  \cmidrule(lr){2-5} \cmidrule(lr){6-10}
        \textbf{Method} & \textbf{MSD} & \textbf{SCE} & \textbf{RMS} & \textbf{LUFS} & & \textbf{MSD} & \textbf{SCE} & \textbf{RMS} & \textbf{LUFS} \\ \midrule
        {Input}   & 10.4        & 1041.7        & 10.4          & 3.6   & & 8.4         & 2607.4      & 9.4        & 3.8   \\ 
        {RB-DSP}  & 8.9         & 517.2         & 5.8           & 2.4   & & 6.5         & \textbf{915.4}       & 8.6         & 3.7   \\  \midrule
        {NP-HH S} & 9.8         & 636.6         & 12.9          & 5.9   & & 7.0         & 1512.7      & 10.0        & 4.5          \\
        {SPSA}    & 8.0         & 360.1         & \textbf{5.1} & 2.5    & & 5.5         & 1297.0      & 4.6         & 2.1         \\
        {AD}      & \textbf{7.8}& \textbf{278.1}& 5.2  & \textbf{2.4}   & & \textbf{4.8} & 947.6       & \textbf{3.8} & \textbf{1.7}             \\ 
         \bottomrule 
    \end{tabular}}
    \vspace{0.2cm}
    \caption{Realistic production style transfer average performance of all pairwise configurations from five predefined styles with speech from DAPS using the model trained on LibriTSS and music from MUSDB18 using the model trained on MTG-Jamendo.}
    \vspace{-0.3cm}
    \label{tab:style-transfer-real}
\end{table}

We performed all pairwise production style transfers resulting in 25 configurations, and we performed each configuration ten times using different input and output recordings from each style. 
Here, we considered only the best performing methods that produce interpretable control parameters (NP-HH, SPSA, and AD).
To summarize the results of this evaluation, we aggregated the metrics across all 25 configurations reporting the mean for each to provide a general notion of the production style transfer performance, as shown in \tref{tab:style-transfer-real}. 

On examples from DAPS, all methods including the baseline improve upon the input, but SPSA and AD clearly outperform the other approaches across all metrics, with AD performing the best. 
In the case of MUSDB18, the AD approach further outperforms the others, with the exception of SCE, where the RB-DSP baseline performs best.


\subsection{Parameter Visualization}
\begin{figure}[!t]
    \centering
    \includegraphics[width=\linewidth, trim={0.3cm 0.1cm 0.85cm 0},clip ]{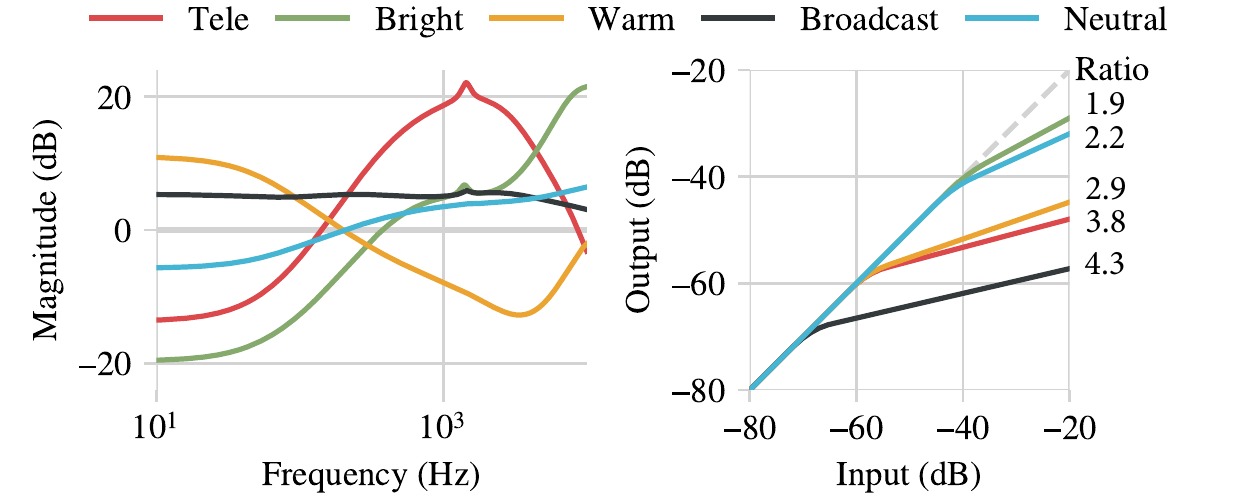}
    \vspace{-0.8cm}
    \caption{Estimated parametric equalizer (left) and compressor (right) parameters from our SPSA model with one utterance from VCTK and five unique style references.}
    \label{fig:vctk-variable-style}
    \vspace{-0.0cm}
\end{figure}

To demonstrate the ability of our approach to provide precise control over the effect parameters, we plot both the overall frequency response of the parametric equalizer as well as the compressor configuration in~\fref{fig:vctk-variable-style}.
We show the result of using one utterance from VCTK as the input with five different style references.
Since the utterances from VCTK are relatively clean and contain little post-processing, we observed the estimated parameters aligned with our expectation for the predefined styles. 
For example, we find that the telephone style is achieved by reducing the high and low frequencies while boosting frequencies around 1\,kHz.
In terms of compression, we see a clear trend with the broadcast style using the most aggressive compression, as expected. 
While estimating these style parameters may seem trivial, our system was not trained on these recordings and has no knowledge of the styles. 

\begin{figure}[!t]
\vspace{-0.0cm}
    \centering
    \includegraphics[width=\linewidth, trim={0.3cm 0.1cm 0.85cm 0},clip ]{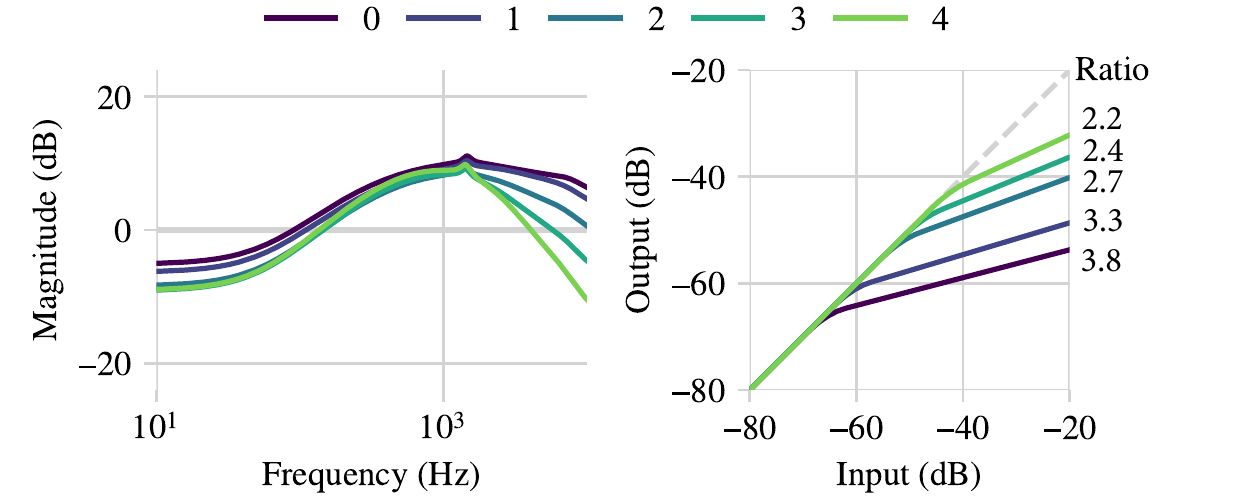}
    \vspace{-0.8cm}
    \caption{Estimated parametric equalizer (left) and compressor (right) parameters with our SPSA model using a fixed Broadcast style reference. The input VCTK utterance was progressively modified to increase the high frequency energy and compression.}
    \label{fig:vctk-variable-input}
    \vspace{-0.2cm}
\end{figure}

As a demonstration of the ability of the model to adapt to the input signal like an \emph{adaptive preset}, we processed a clean recording using a reference from the Broadcast style while varying characteristics of the input. 
In this case, the input signal was modified by progressively increasing the gain of a high-shelf filter from 0\,dB to +24\,dB, while also increasing the amount of compression with a higher ratio (1.0 to 4.0) and lower threshold (0\,dB to -62\,dB).
As shown in \fref{fig:vctk-variable-input}, the estimated parameters adapt to these changes in a sensible manner. 
This results in a progressively lower boost in the high-shelf filter that was added originally to achieve the presence of the Broadcast style, while also increasing the threshold and decreasing the ratio to reduce the amount of compression, as the input becomes more compressed. 

\subsection{Encoder Representations}
\setlength{\tabcolsep}{3.0pt}

\begin{table}[t!]
    \centering
    \resizebox{1.0\hsize}{!}{%
    \begin{tabular}{l c c c c c c} \toprule
        \multicolumn{7}{c}{\textbf{DAPS} (Speech)} \\ \midrule
        \textbf{Features} & Telephone &  Bright & Warm & Broadcast & Neutral & \textbf{Avg}\\  \midrule
        Random Mel      & 0.87 & 0.78 & 0.73 & 0.39 & 0.00 & 0.55\\
        OpenL3          & 0.19 & 0.61 & 0.08 & 0.10 & 0.18 & 0.23\\ 
        CDPAM           & \textbf{1.00} & \textbf{1.00} & 0.79 & 0.25 & 0.63 & 0.73\\\midrule
        NP-HH S         & \textbf{1.00} & \textbf{1.00} & \textbf{1.00} & \textbf{1.00} & \textbf{1.00} & \textbf{1.00}\\
        SPSA            & 0.95 & 0.98 & \textbf{1.00} & 0.89 & 0.95 & 0.96\\
        AD              & \textbf{1.00} & \textbf{1.00} & \textbf{1.00} & \textbf{1.00} & \textbf{1.00} & \textbf{1.00}\\ \midrule
        \multicolumn{7}{c}{\textbf{MUSDB18} (Music)} \\ \midrule
        \textbf{Features} & Telephone &  Bright & Warm & Broadcast & Neutral & \textbf{Avg} \\  \midrule
        Random Mel        & 0.80 & 0.98 & 0.62 & 0.17 & 0.00 & 0.51 \\
        OpenL3            & 0.32 & 0.66 & 0.20 & 0.17 & 0.30 & 0.33 \\ 
        CDPAM             & 0.89 & 0.95 & 0.66 & 0.00 & 0.06 & 0.51 \\\midrule
        NP-HH S         & \textbf{0.98} & \textbf{1.00} & 0.92 & \textbf{0.59} & \textbf{0.60} & \textbf{ 0.82} \\
        SPSA            & \textbf{0.98} & \textbf{1.00} & 0.90 & 0.26 & 0.00 & 0.63 \\
        AD              & \textbf{0.98} & \textbf{1.00} & \textbf{0.95} & 0.54 & 0.50 & 0.79 \\
         \bottomrule 
    \end{tabular}}
    \vspace{0.2cm}
    \caption{Class-wise F1 scores for five-class style prediction with linear classifiers trained on top of audio representations for speech and music using a single linear layer.}
    \label{tab:style-probes}\    
    \vspace{-0.2cm}
\end{table}

In addition to our production style transfer evaluation, we also investigated the potential utility of the representations learned by our encoder. 
To do so, we created a downstream task of audio production style classification, where the goal is to classify the style of a recording among the five styles we created for our realistic style evaluation. 
We trained linear logistic regression classifiers, also known as linear probes, on top of a set of different audio representations, a common approach in evaluating the quality of self-supervised representations~\cite{zhang2017split, asano2019critical}.
These linear probes are trained for a total of 400 epochs on a small dataset with only 60 examples from each of the five styles.
We show class-wise and overall F1 scores on the test set in Table~\ref{tab:style-probes}.

As baseline representations, we considered a random projection of melspectrogram features, along with large pre-trained representations including OpenL3~\cite{cramer2019look}, which is focused on capturing audio content and CDPAM~\cite{manocha2021cdpam}, which is focused on capturing audio quality, therefore the most similar to our approach.
In addition, we trained linear probes on representations from the NP-HH, SPSA, and AD differentiation methods on both speech styles from DAPS and music styles from MUSDB18.

In the case of DAPS, the class-wise scores indicated that CDPAM and our proposed representations were able to capture the Bright and Telephone styles. 
Classifying between Broadcast and Neutral was difficult due to the main difference between these styles being the presence of more aggressive compression. 
However, we found that our encoder representations did not struggle to capture this difference. 
Overall, it is clear that representations from our approach not only capture style elements related to spectral content, but also signal dynamics, which are not captured well by the other approaches. 
We found similar results for both speech and music, where music was slightly more difficult, likely due to MUSDB18 style diversity.


\subsection{Computational Complexity}
Beyond production style transfer performance, we benchmark training time per step, CPU/GPU inference time, and number of parameters for audio effects per method variant in~\fref{tab:runtime}.
To compare the training step we measure the time required for both a forward and backward pass when optimizing the system using the same batch size and input size as training.
We found the neural proxy and TCN approaches impart the highest computational cost since they feature the use of neural networks for audio processing during both inference and training. 
However, the neural proxy hybrids (HP-HH and NP-FH) and SPSA have the lowest computational cost since they all use only the DSP device during inference, which is on par with RB-DSP baseline. 
The AD approach is twice as slow as the DSP-based methods, but is still far more efficient than any of the neural network based effects. 

\setlength{\tabcolsep}{3.0pt}
\begin{table}[t]
    \centering
    \resizebox{1.0\hsize}{!}{%
    \begin{tabular}{l c c c c c} \toprule
        \textbf{Method}  &  \textbf{Train step (s)}   &  \textbf{CPU} &  \textbf{GPU} & \textbf{Parameters} & \textbf{Interpretable}\\ \midrule
        RB-DSP      & -     & 0.004 & -     & 0 & $\mathcal{-}$\\ \midrule
        cTCN 1      & 0.438 & 0.132 & 0.002 & 174\,k & - \\
        cTCN 2      & 0.642 & 0.268 & 0.005 & 336\,k & - \\ \midrule
        NP          & 0.434 & 0.277 & 0.005 & 336\,k & \checkmark \\
        NP-HH       & 0.434 & \textbf{0.003} & -     & 0 & \checkmark\\
        NP-FH       & 0.434 & \textbf{0.003} & -     & 0 & \checkmark \\
        SPSA        & 0.413 & \textbf{0.003} & -     & 0 & \checkmark \\
        AD          & \textbf{0.301} & 0.006 & \textbf{0.001} & 0 & \checkmark \\
                 \bottomrule 
    \end{tabular}}
    \vspace{0.2cm}
    \caption{Runtime comparison across differentiation methods including seconds taken for a single training step, and real-time factor for inference on CPU (Intel Xeon CPU E5-2623 v3 @ 3.00GHz) and GPU (GeForce GTX 1080 Ti). }
    \label{tab:runtime}
     \vspace{-0.2cm}
\end{table}

\section{DISCUSSION}
\label{sec:discussion}

Our production style transfer approach provides multiple  user-interaction paradigms. 
At a high-level, a user can supply their own input content and style reference and view the estimated control parameters. 
This affords significant agency, but requires the user have a goal and a suitable reference recording. 
We can simplify this interaction by creating \emph{adaptive presets} that supply a curated, predefined set of style reference recordings, each with semantic descriptions. 
Lastly, we can also empower a fully automatic solution by employing a single, predefined style reference.
 
When we compare differentiation methods, we note the AD approach performed best across most metrics. There are, however, a number of additional considerations. 
In particular, the AD approach is the most manual and demanding method to implement, commonly requires significant expertise and design trade-offs to implement, and can be more difficult to integrate into on-device inference. 
In contrast, generic neural network architectures like our TCN-based baselines performed well in comparison and are ultimately more flexible. 
However, these approaches do not provide interpretable control and require significant computation, making them undesirable and difficult to integrate into audio production environments. 

Lastly, when we compare against the SPSA and NP methods, we note they offer interpretable control (similar to AD), but do not require knowledge or re-implementation of the underlying system (black-box), do not require explicit differentiation, enable the use of the original DSP at inference, and typically require less computation. 
We observed, however, NP and SPSA methods have other drawbacks.
We found SPSA was more susceptible to training instability and required careful tuning of the $\epsilon$ hyperparameter and learning rate. 
And, while the NP variants were promising, we found they were unable to achieve performance on par with the other approaches. We hypothesize this was caused by inaccuracy in the proxy networks, especially for the parametric equalizer. 
Strict NP approaches also require laborious pre-training and design.

We conclude that the preferred differentiation strategy for our task and similar automatic production methods is application dependent.  
Future research directions include the design of better neural proxy methods, a reusable differentiable audio effects library, the ability for dynamic audio signal chain construction, and/or improved numerical gradient approximation methods.

\section{CONCLUSIONS}
Our \emph{DeepAFx-ST} method imposes the production style from one recording to another, enabling automatic audio production and \emph{adaptive presets}.
We demonstrated the applicability of our approach in both speech and music tasks and performed an extensive comparison of a diverse range of differentiable signal processing methods. 
We proposed a categorization of existing approaches for differentiable signal processing, and also introduced differentiable dynamic range compressor and neural proxy hybrid approaches.
Our approach demonstrated convincing performance on both the synthetic and realistic production style transfer tasks, while estimating interpretable audio effect controls, operating at variable sample rates, and producing representations that accurately reflect production style.

\section{ACKNOWLEDGMENT}
This work is supported in part by the EPSRC UKRI CDT in Artificial Intelligence and Music (EP/S022694/1). We would like to thank O. Wang, P. Smaragdis, and J.-P. Caceres for their valuable conversations.

\bibliography{aes2e.bib}
\bibliographystyle{aes2e.bst}

\biography{Christian J. Steinmetz}{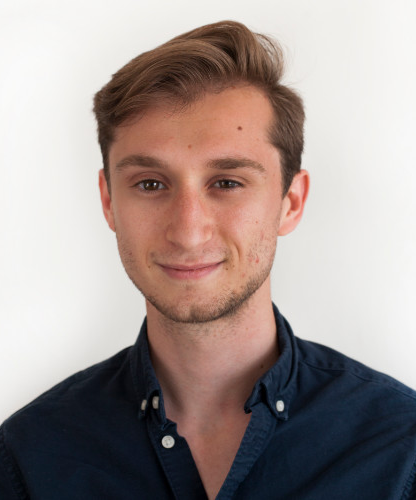}{Christian J. Steinmetz is a PhD researcher with the Centre for Digital Music at Queen Mary University of London. His research focuses on applications of machine learning for audio signal processing with a focus on high fidelity audio and music production. This involves building tools for enhancing audio recordings, automatic and assistive systems for audio engineering, as well as applications of machine learning that augment creativity in music production. He recently won the Best Student Paper Award at the IEEE Workshop on Applications of Signal Processing to Audio and Acoustics for his work in blind room impulse response estimation. He has worked as a research scientist intern at Adobe, Facebook Reality Labs, and Dolby Labs. Christian holds a BS in Electrical Engineering and BA in Audio Technology from Clemson University, as well as an MSc in Sound and Music Computing from the Music Technology Group at Universitat Pompeu Fabra.}

\biography{Nicholas J. Bryan}{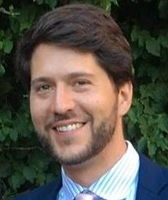}{Nicholas J. Bryan is a Senior Research Scientist at Adobe Research, San Francisco. His research focuses on machine learning and signal processing. Nick has published over 32 peer reviewed papers, 8 patents (+9 under review), and won 2 best paper awards, 1 AES Graduate Student Design Gold Award, and 1 best reviewer award. He is a member of the IEEE Audio and Acoustic Signal Processing Technical Committee. Nick received his PhD and MA from CCRMA, Stanford University and MS in Electrical Eng., also from Stanford, and his BM and BS in Electrical Eng. with summa cum laude, general, and departmental honors at the U. of Miami-FL. Before Adobe, Nick was a senior audio algorithm engineer at Apple.} 

\biography{Joshua D. Reiss}{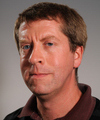}{Josh Reiss is Professor of Audio Engineering with the Centre for Digital Music at Queen Mary University of London. He has published more than 200 scientific papers (including over 50 in premier journals and 6 best paper awards) and co-authored two books. His research has been featured in dozens of original articles and interviews on TV, radio, and in the press. He is a Fellow and currently President of the Audio Engineering Society (AES), and chair of their Publications Policy Committee. He co-founded the highly successful spin-out company, LandR, and recently co-founded Tonz and Nemisindo, also based on his team’s research. He maintains a popular blog, YouTube channel, and Twitter feed for scientific education and dissemination of research activities.}
\end{document}